\documentclass[a4paper]{jpconf}
\usepackage{graphicx}
\begin{document}
\title{SpectroWeb: oscillator strength measurements of atomic absorption lines in the Sun and Procyon}

\author{A. Lobel}

\address{Royal Observatory of Belgium, Ringlaan 3, B-1180, Brussels, Belgium}

\ead{alobel@sdf.lonestar.org}

\begin{abstract}
We update the online SpectroWeb database of spectral standard  
reference stars with 1178 oscillator strength values of atomic 
absorption lines observed in the optical spectrum of the 
Sun and Procyon ($\alpha$ CMi A). 
The updated line oscillator strengths are measured
with best fits to the disk-integrated KPNO-FTS  
spectrum of the Sun observed between 4000 \AA \, and 6800 \AA\
using state-of-the-art detailed spectral synthesis calculations.
A subset of 660 line oscillator strengths is validated with 
synthetic spectrum calculations of Procyon observed with 
ESO-UVES between 4700 \AA\, and 6800 \AA. The new log(gf)-values 
in SpectroWeb are improved over the values
offered in the online Vienna Atomic Line Database (VALD). 
We find for neutral iron-group elements, such as 
Fe~{\sc i}, Ni~{\sc i}, Cr~{\sc i}, 
and Ti~{\sc i}, a statistically significant over-estimation of the 
VALD log(gf)-values for weak absorption lines with normalized 
central line depths below 15~\%. For abundant lighter elements 
(e.g. Mg~{\sc i} and Ca~{\sc i})
this trend is statistically not significantly detectable, with 
the exception of Si~{\sc i} for which the log(gf)-values of 60 weak and 
medium-strong lines are substantially decreased to best fit the 
observed spectra. The newly measured log(gf)-values are available 
in the SpectroWeb database at \verb"http://spectra.freeshell.org"
which interactively displays the observed and computed stellar 
spectra, together with corresponding atomic line data.   
\end{abstract}

\begin{figure}[h]
\vspace*{-3cm}
\includegraphics[width=3.7in]{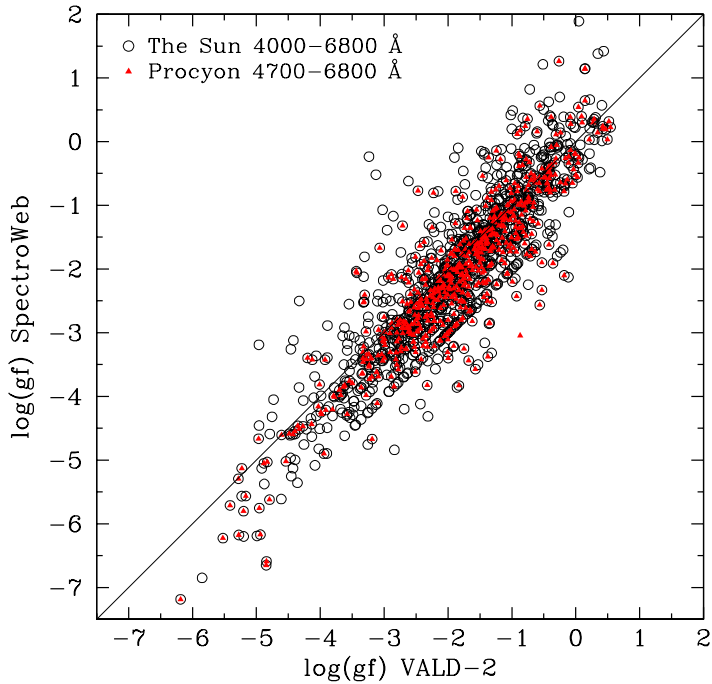}
\hspace{1pc}
\begin{minipage}[b]{14pc}
\caption{\label{label}Log(gf)-values of 1178 absorption lines that 
best fit the solar optical spectrum in SpectroWeb 
compared to the VALD values ({\em open symbols}). 
Filled triangles mark a subset of 660 lines verified in Procyon. 
The majority of VALD log(gf)-values  (mostly from weak lines of iron-group 
elements) are over-estimated compared to the values measured in SpectroWeb.}
\end{minipage} 
\end{figure}

\section{Introduction}

The SpectroWeb database is an online repository of identified spectral 
lines and features observed in spectral standard reference stars. 
It is permanently updated and improved, currently providing 
high-resolution spectra of six bright (cool) stars selected 
as primary  spectroscopic reference objects: 
Betelgeuse ($\alpha$ Ori; M2 Iab),
Arcturus ($\alpha$ Boo; K1 III), The Sun (G2 V), $\beta$ Aqr (G0 Ib),
Procyon ($\alpha$ CMi A; F5 IV-V), and Canopus ($\alpha$ Car; F0 II).
Their effective temperatures differ by about 1000 K, ranging from 
3500 K (M-type) to 7500 K (F-type). These stars offer a broad range 
of thermal conditions for the identification of mainly neutral and singly 
ionized spectral lines formed in their atmospheres. SpectroWeb offers a comprehensive interactive database of identified spectral 
lines that relies on detailed comparisons of observed spectra with advanced spectrum synthesis calculations. With its graphics display 
users can zoom in on the same wavelength regions of interest in different stars to investigate changes of line intensities, and to  
directly assess the reliability of the line identifications and the
quality of the corresponding atomic line data. SpectroWeb is freely
accessible online at \verb"spectra.freeshell.org". The database's graphics interface requires a modern internet browser with an activated Java language interpreter. The object-oriented (Java `applet') implementation, 
for example, permits to securely link many digital spectral atlases in 
a single database that is served from various world-wide-web domains 
using a standard interactive display. 
A concise description of the current SpectroWeb 1.0 implementation 
and its basic query interactions is provided in \cite{lobelnum1}.

\section{Observed and theoretical spectra in SpectroWeb}

The high-resolution spectrum of the Sun observed with the 
NSO/KPNO Fourier Transform Spectrograph (FTS) is offered in 
\cite{neckelnum1}. The 
Procyon spectrum observed with the ESO Ultraviolet and Visual 
Echelle Spectrograph (UVES) is offered in the ESO Science Archive 
\cite{bagnulonum1}. 
The spectral resolving power $R$ of the disk integrated 
FTS spectrum is $\sim$350,000, while the nominal UVES resolution 
is $\sim$80,000. The S/N ratio of the Procyon echelle spectrum 
of October 2002 is 300 to 500 in the $V$-band, which is sufficiently
large to resolve weak absorption features with central depths 
exceeding 2~\% of the normalized stellar continuum flux level. 
More information about the VLT-UVES instrument and pipeline calibration 
is given in \cite{dekkernum1} and \cite{bagnulonum1}. The S/N ratio of the 
FTS solar spectrum observed in 1981 at the KPNO-McMath-Pierce Solar 
Facility is estimated around 2,500. More information about the
calibration of the solar mean intensity atlas is provided in \cite{neckelnum2}.
The spectra of Betelgeuse, Arcturus, and Canopus are also obtained 
from the ESO-UVES Archive, while the $\beta$ Aqr spectrum is 
from the Elodie Archive at the OHP. A discussion of the 
latter four spectra will be given elsewhere. It is more 
important to point out that the increase of about 1000 K 
between these stars towards earlier spectral types yields
strong changes in the optical spectrum due to large changes 
of the stellar atmospheric ionization balance. 
For the coolest stars line blending strongly increases towards 
shorter wavelengths, resulting in a large decrease of the local 
continuum  flux level to below the stellar continuum level. 
For M-supergiant Betelgeuse the optical spectrum is 
dominated by molecular opacity and mainly due to TiO.

\begin{figure}[h]
\vspace*{-3cm}
\begin{minipage}{17pc}
\includegraphics[width=3.in]{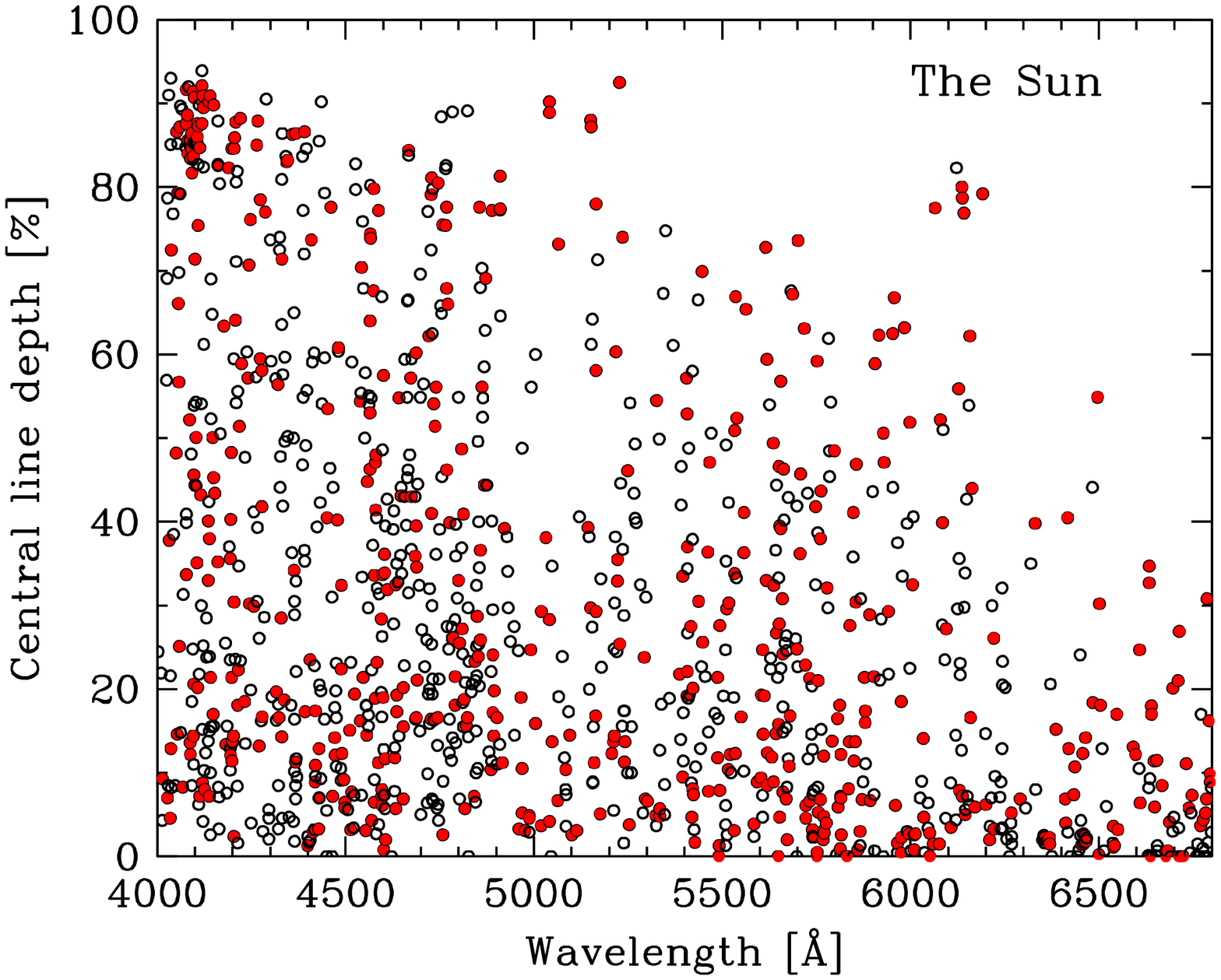}
\caption{\label{label}Computed normalized central line depths of 
1178 atomic absorption lines in the Sun of which the log(gf)-values
are measured to fit the observed spectrum. Filled symbols mark Fe~{\sc i} lines.}
\end{minipage}\hspace{3pc}%
\begin{minipage}{17pc}
\includegraphics[width=3.in]{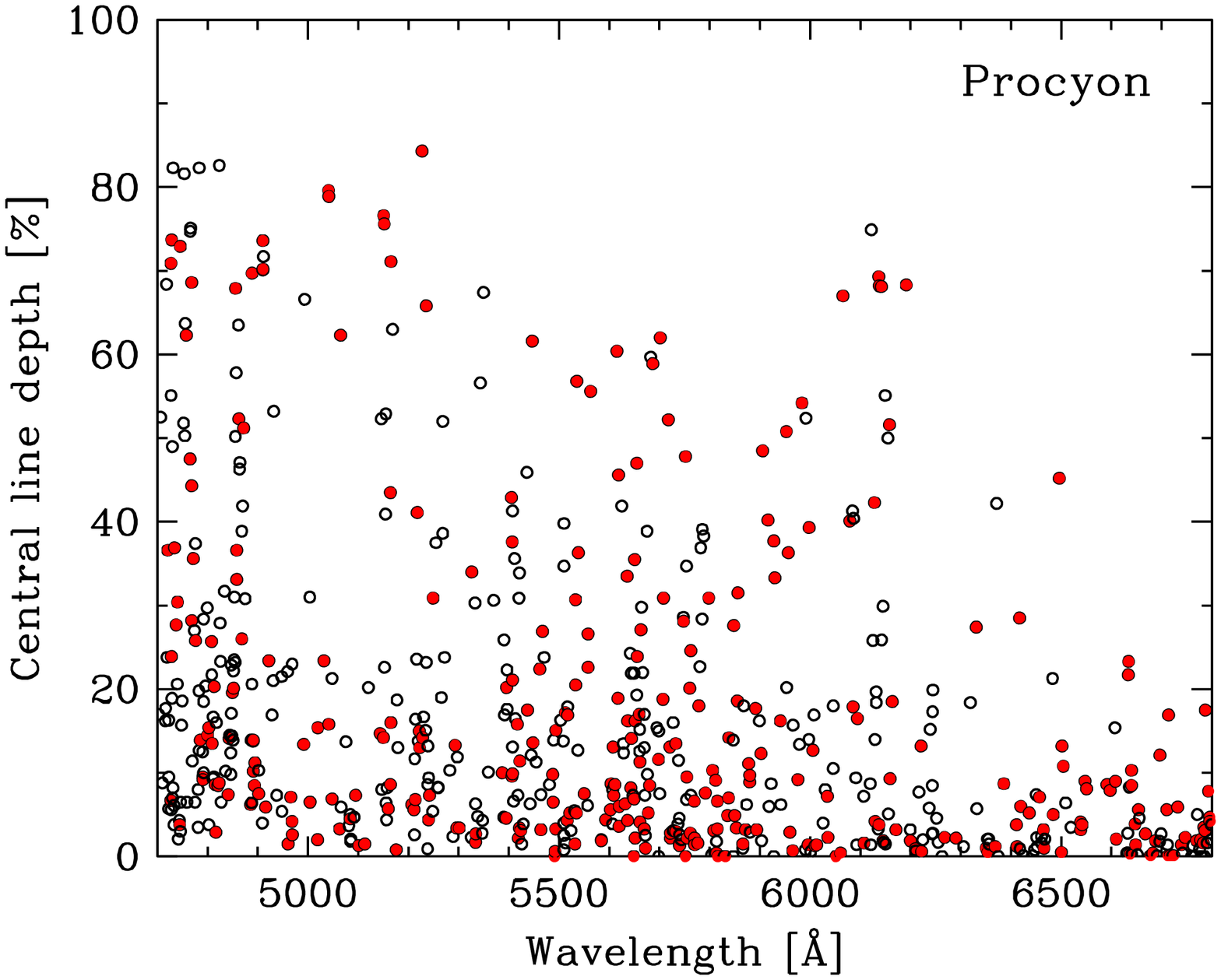}
\caption{\label{label}Same as Fig. 2 for a subset of 660 lines in Procyon between 4700 \AA\, and 6800 \AA. The Fe~{\sc i} lines and other
neutral lines become weaker in Procyon due to its larger $T_{\rm eff}$. }
\end{minipage} 
\end{figure}

The placement of the stellar continuum flux level to provide 
the continuum normalized echelle atlases in SpectroWeb is 
based on detailed spectral synthesis calculations between 
3300 \AA\, and 6800 \AA. The observed spectra are  
converted to the stellar rest wavelength scale to facilitate 
an accurate comparison to the theoretical spectra. 
The latter spectra are computed with radiative transfer 
in LTE using 1D hydrostatic models of the stellar atmosphere. 
The atmospheric parameters in the model grid \cite{kurucznum1}
$T_{\rm eff}$, log $g$, and the projected microturbulence velocity 
are varied until an overall best fit to the observed spectrum 
is obtained. It involves an iterative fit procedure whereby
the differences of relative line depths between the observed 
and computed high-resolution spectra are minimized. 
The fit procedure utilizes the `normal' stellar photospheric 
spectrum in wavelengths regions that are void of blends 
with strong telluric lines. It also excludes the broad 
H~{\sc i} Balmer lines, and the strong 
doublet lines of Ca~{\sc ii} H \& K and Na~{\sc i} $D$. The 
current atmosphere models omit chromospheric structures 
for these cool stars (including the Sun) which can alter the 
depth and detailed shape of the resonance lines. 
The detailed modeling of these broad line profiles requires 
semi-empiric radiative transfer calculations in non-LTE \cite{lobelnum2}.

\begin{figure}[h]
\vspace*{-1cm}
\begin{minipage}{17pc}
\includegraphics[width=2.8in]{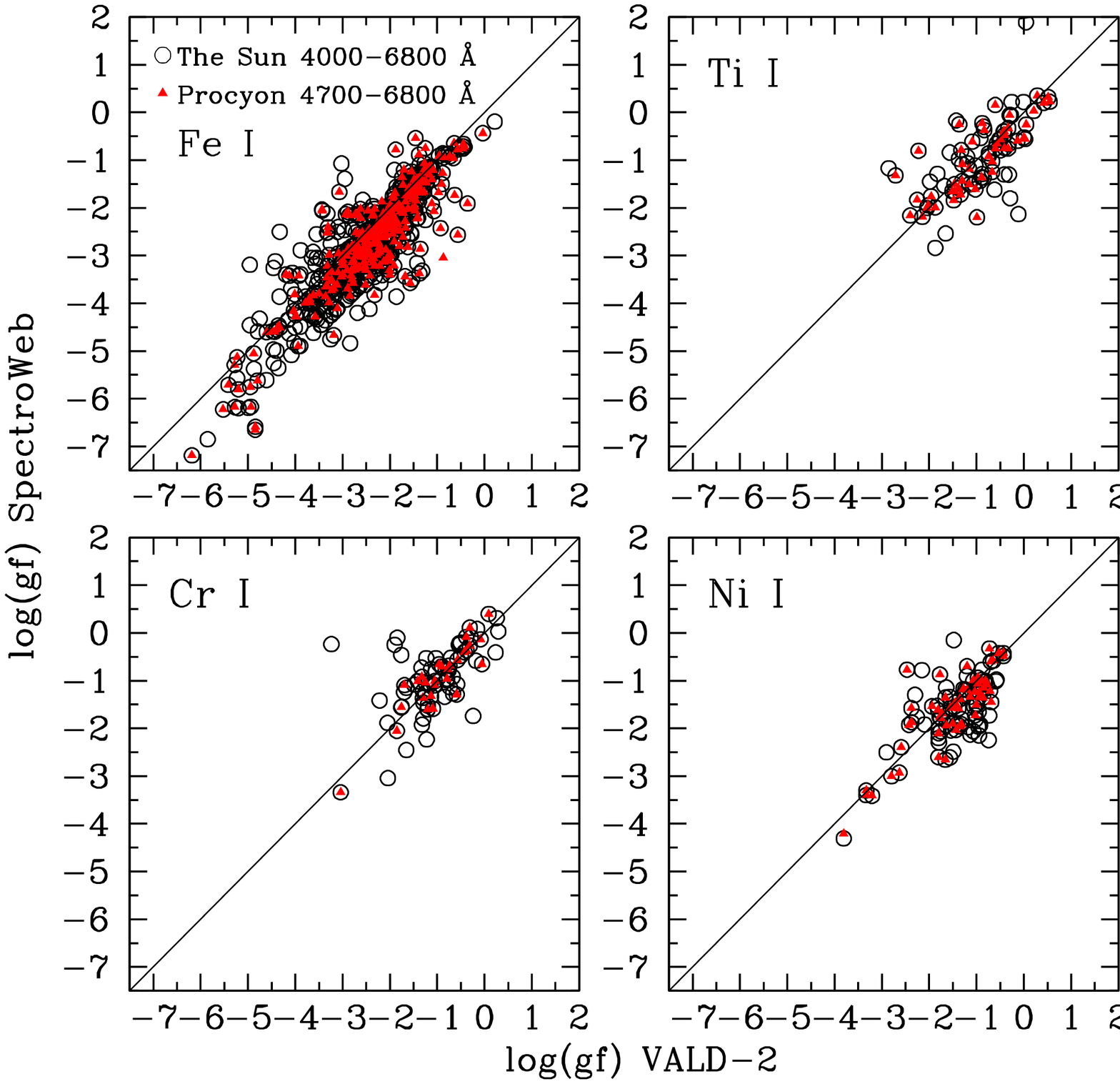}
\caption{\label{label}Same as Fig. 1 shown for neutral lines of Fe, Ti, Ni, and Cr ({\em clockwise}). More neutral lines of Fe, covering a broader 
range of log(gf)-values, are selected in the optical spectrum for detailed 
spectral synthesis compared to Ti, Ni, and Cr. } 
\end{minipage}\hspace{3pc}%
\begin{minipage}{17pc}
\includegraphics[width=2.8in]{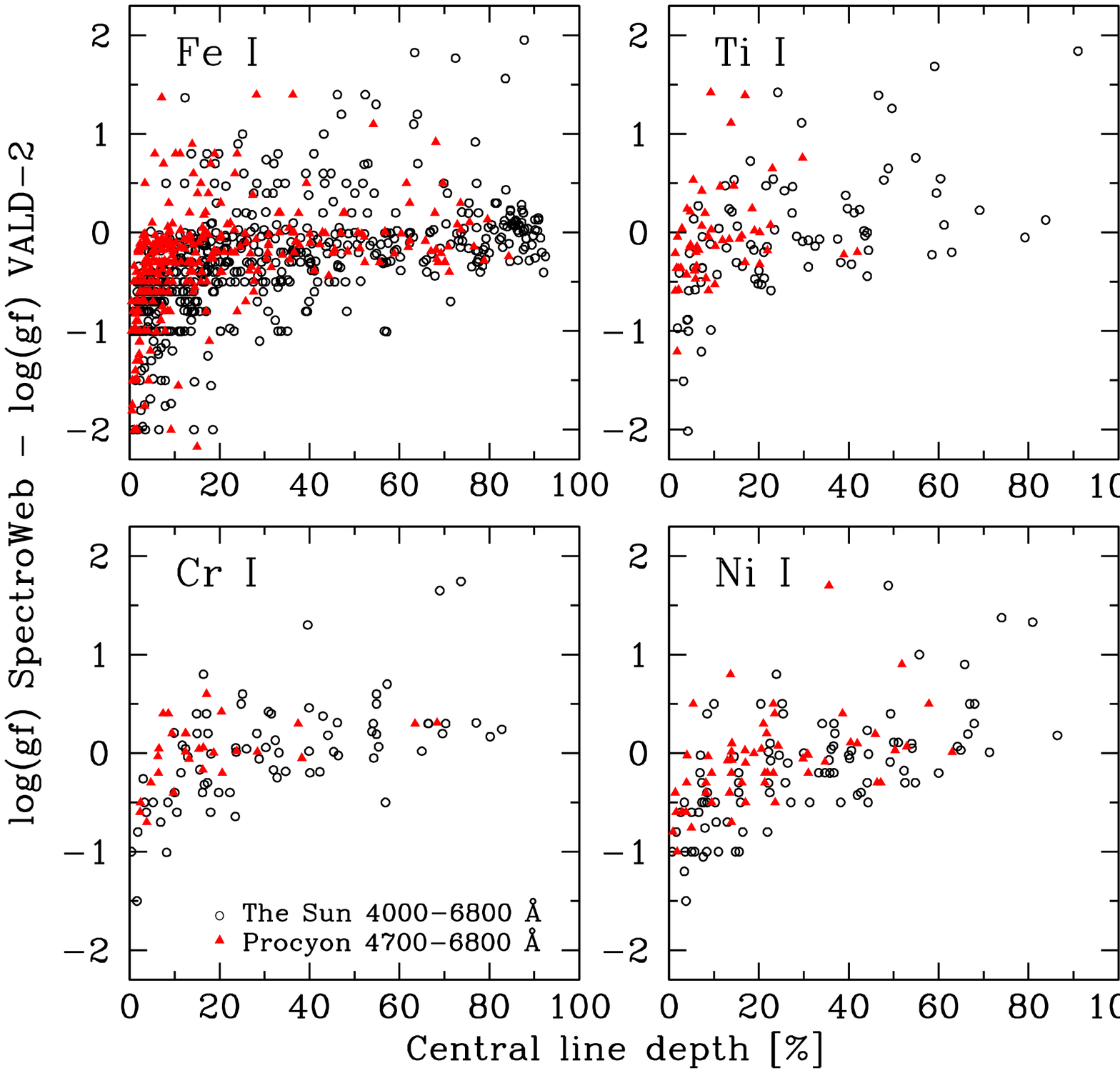}
\caption{\label{label}Differences between log(gf)-values measured in SpectroWeb and offered in VALD compared to the normalized central depth 
computed for neutral lines of four iron-group elements in Fig. 4. The log(gf)-values of weak lines are systematically over-estimated in VALD.}
\end{minipage} 
\end{figure}

The input lists of spectral lines for radiative transfer 
are obtained from \cite{kurucznum2}. 
The detailed atomic line data 
in SpectroWeb is adopted from the online Vienna Atomic Line  
Database (VALD) \cite{kupkanum1}, providing values of the line 
oscillator strength 
(log(gf)), the transition energy levels, together with 
the other line broadening parameters. A large number of 
diatomic molecular lines is incorporated 
to improve the position of the stellar continuum level.
The spectra are currently computed for solar elemental abundance 
values with models of [M/H]=0.0. We use the elemental abundance 
values of \cite{andersnum1} for consistency with our atmospheric models. 
For example, we currently adopt [Fe/H]=7.67 instead 
of the more recent and appreciably smaller (meteoritic) value 
of 7.5 \cite{grevessenum1}. The theoretical spectra do currently 
not include telluric lines due to water vapor and $\rm O_{2}$ in Earth's 
atmosphere. The position of the strongest $\rm H_{2}O$ and 
$\rm O_{2}$ lines are only marked. The spectra are convolved 
with a filter to simulate the instrumental profile 
of the observed spectra.

\section{Line oscillator strength measurements in the Sun and Procyon} 
We find best spectrum fits using a constant microturbulence velocity 
of 1.1 $\rm km\,s^{-1}$ in the solar atmosphere model and 1.2 $\rm km\,s^{-1}$ in Procyon. The synthetic spectra are rotationally convolved with $v$sin$i$ values of 2.5 $\rm km\,s^{-1}$
and 3.6 $\rm km\,s^{-1}$, respectively. Hyperfine line splitting 
has not been incorporated so far. Figure 1 shows 1178 lines 
for which we correct the VALD-2 log(gf)-values to the
values in SpectroWeb, yielding the best fit 
to the solar spectrum between 4000 \AA\, and 6800 \AA\, 
({\it open symbols}). The corrected log(gf)-values of 660 lines 
between 4700\AA\, and 6800 \AA\, have currently been   
validated against the spectrum of Procyon ({\it filled triangles}).
The amount of corrected weak, medium-strong, and strong 
absorption lines is almost uniformally distributed over both 
wavelength bands. Figures 2 \& 3 indicate a somewhat smaller 
number of strong lines that are corrected longward 
of $\sim$6000 \AA\, because the total number of strong 
lines diminishes longwards in both spectra. The amounts of 
corrected Fe~{\sc i} lines ({\it filled symbols}) 
with central normalized depths below 40~\% (we compute 
without instrumental broadening) are 
uniformally distributed in both stars. The neutral 
lines become weaker in Procyon because 
$T_{\rm eff}$ is $\sim$1000 K larger
and we can also adopt solar abundance values \cite{katonum1}.

\begin{figure}[h]
\vspace*{-1cm}
\begin{minipage}{17pc}
\includegraphics[width=2.4in]{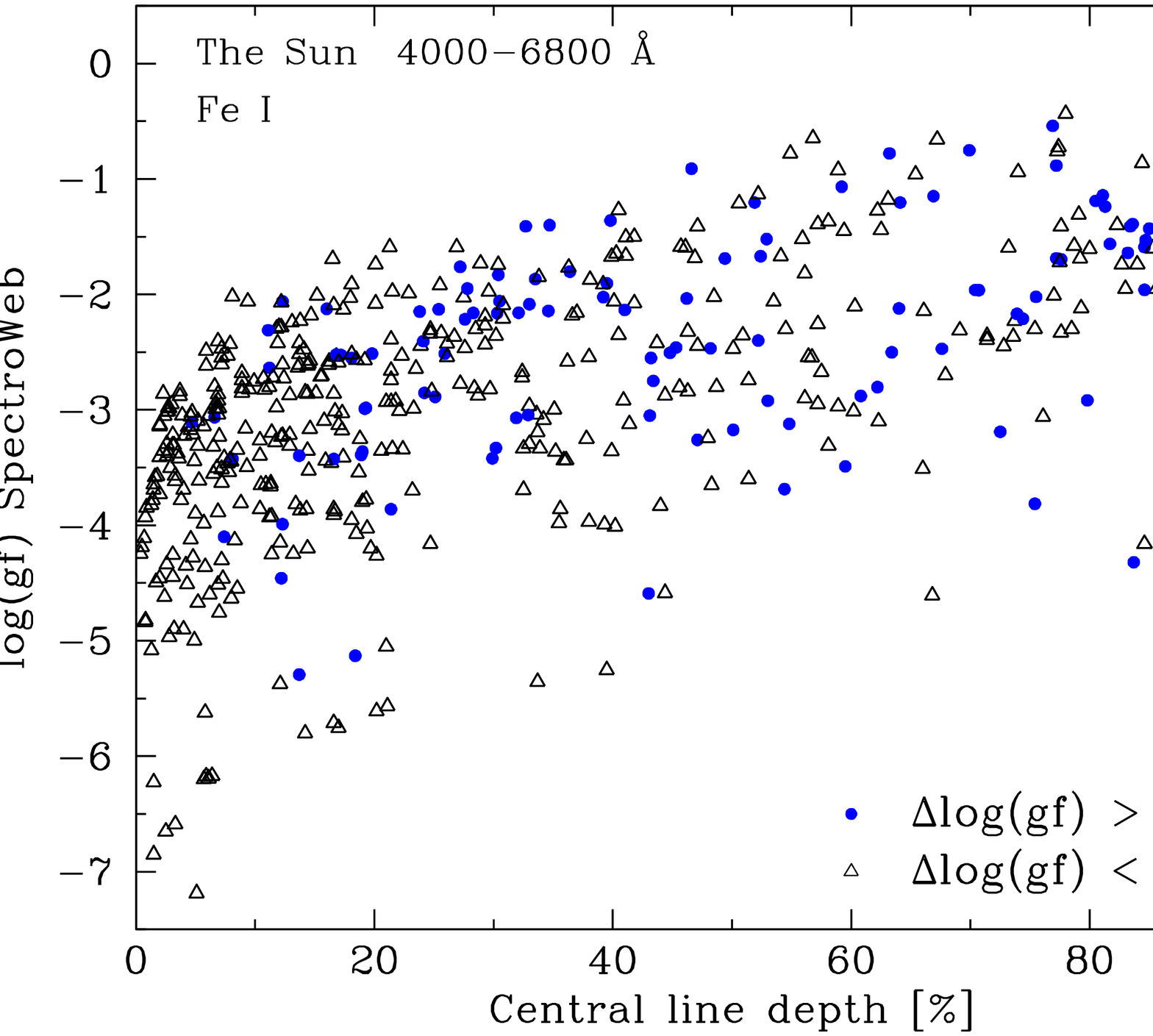}
\caption{\label{label}Log(gf)-values of Fe~{\sc i} 
lines in SpectroWeb measured from detailed spectral synthesis fits 
to the solar optical spectrum against the computed central line depth.
Open symbols mark lines with decreased log(gf)-values compared to 
VALD, while filled symbols mark increased values. Significantly more
weak lines require a log(gf) decrease than an increase.   
}\end{minipage}\hspace{3pc}%
\begin{minipage}{17pc}
\includegraphics[width=2.4in]{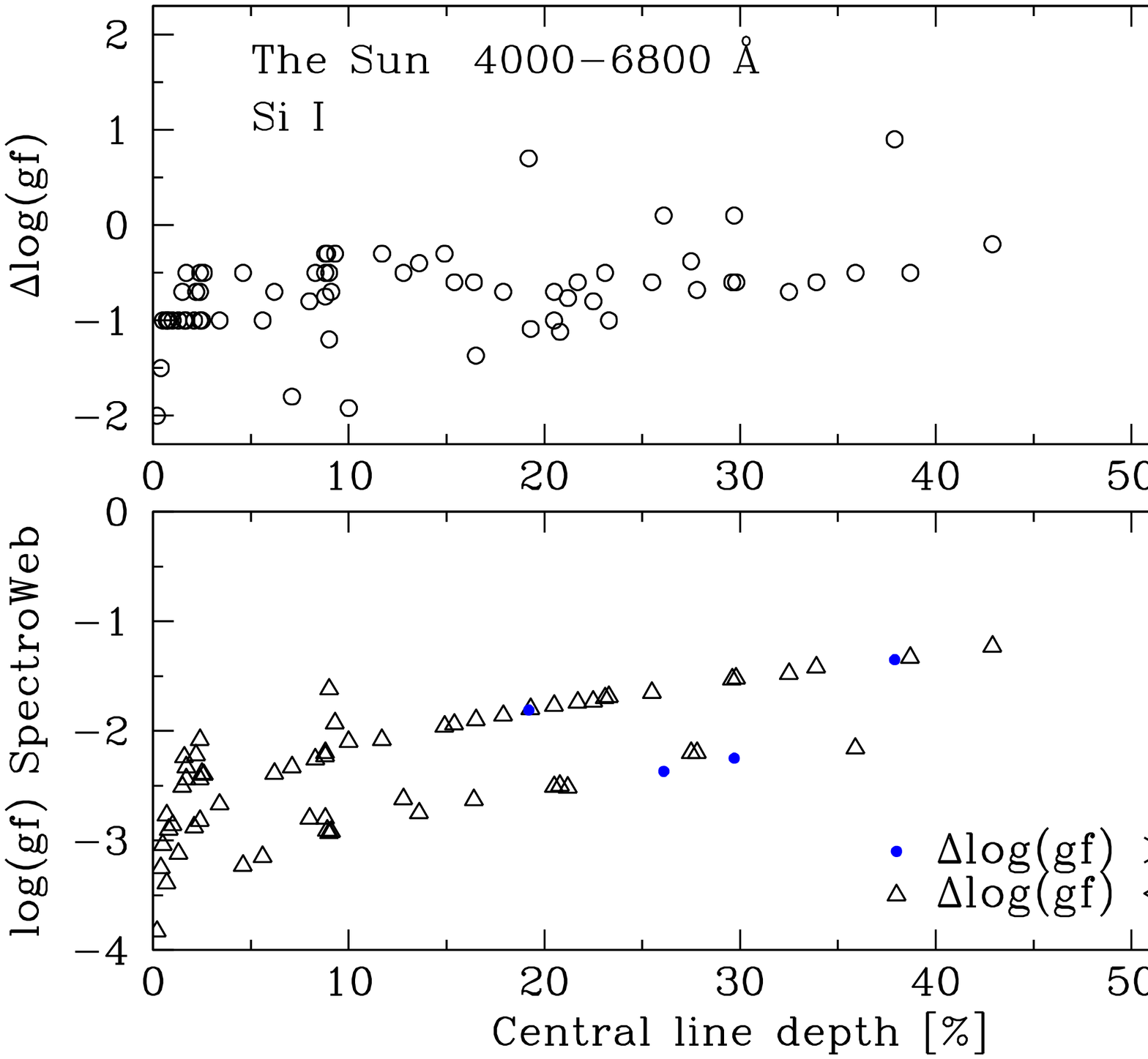}
\caption{\label{label}Upper panel: the log(gf)-values of 60 weak and medium-strong Si~{\sc i} lines are over-estimated in VALD, while only 5 lines require a log(gf) increase. Lower panel: there are multiplet dependences between measured log(gf)-values and computed central depths for weak and medium-strong lines in the sample of corrected Si~{\sc i} lines ({\it see text}). }
\end{minipage} 
\end{figure}

Figure~1 reveals that the majority of lines we correct 
 in both stars require log(gf)-values appreciably smaller
than the VALD values. The 
center-of-gravity of the point cloud of measured lines in Fig.~1 
is located below the diagonal line ({\it thin drawn line}) 
around ($-$2.0, $-$2.5) in the linear logarithmic scale.
Figure 4 shows comparable plots of corrected
log(gf)-values in Fig. 1 for individual lines of 
Fe~{\sc i}, Ti~{\sc i}, Ni~{\sc i}, and Cr~{\sc i}. 
We find a statistically significant over-estimation of 
the VALD log(gf)-values in our sample of measured  
Fe~{\sc i} lines. We also measure an over-estimate of
the VALD log(gf)-values in our sample of
Ni~{\sc i} lines, although it covers a considerably 
smaller range of log(gf)-values than the Fe~{\sc i} sample. 
We find that the log(gf)-values of primarily weak 
lines, with normalized central line depths below 15~\%,
are on average over-estimated in the VALD database. 
Figure~5 shows a plot of the lines in Fig. 4 with the measured log(gf)-correction ($\Delta$log(gf) equals the log(gf)-value of SpectroWeb minus the VALD log(gf)-value) compared to 
the normalized line depth.
We observe the trend of over-estimated VALD log(gf)-values in weak 
lines of all iron-group elements for which we measure
a sufficiently large number of lines. On the other hand, 
for the medium-strong and strong lines (e.g. with 
line depths $\geq$ 20~\%) our log(gf)-measurements yield almost equal amounts of lines for which the VALD values are over- and under-estimated. 
In Fig. 6 the VALD log(gf)-values of weak Fe~{\sc i} lines observed 
in the Sun are almost systematically decreased ($\Delta$log(gf)$<$0) to log(gf)-values in SpectroWeb below $-$2.0. The stronger lines require 
$\Delta$log(gf)$<$0  ({\it open triangles}) and $\Delta$log(gf)$>$0 ({\it filled symbols}) corrections for comparable amounts of lines 
to best fit the observed solar spectrum.
A similar trend of over-estimated log(gf)-values from VALD 
for weak lines of lighter elements ($Z$$<$21) is 
statistically not significantly detectable because 
our sample of lines is too small. We find this 
with the exception of the neutral lines of $\alpha$-element 
Si in the Sun. For 60 weak {\em and} medium-strong Si~{\sc i} 
lines in a sample of 65 Si~{\sc i} lines (Fig. 7) we 
measure $\Delta$log(gf)$<$0 ({\it upper panel}). Although the
number of 65 Si~{\sc i} lines is rather limited
we find that the log(gf)-values of both weak {\em and} 
medium-strong lines are significantly over-estimated in VALD
because there are tight dependences between the 
corrected log(gf)-values and the computed line depths  
for Si~{\sc i} lines belonging to the same multiplets. 
The lower panel of Fig. 7 shows these dependences which
are related to the curve-of-growth of the line equivalent 
widths.

Possible shortcomings in our atmospheric model structures 
or spectral synthesis calculations cannot readily explain
a systematic over-estimation of the VALD log(gf)-values 
for weak neutral lines of iron-group elements in 
the Sun and Procyon. Important systematic 
effects due to non-LTE and the chromosphere are expected 
for strong lines rather than for weak lines. A decrease 
of the model abundances of all iron-group elements
would systematically decrease the over-estimated line depths 
we compute with the VALD log(gf)-values for weak lines 
in our sample. However, lowering the abundances of all iron-group 
elements is not an option since a decrease of $\sim$0.2 dex is 
currently adopted for Fe only \cite{grevessenum1}. An abundance decrease 
for iron-group elements would also offset numerous 
lines in our theoretical spectra that do correctly match 
the observed spectra and that are exempt from our 
sample of 1178 corrected lines. We also perform 
spectral synthesis calculations with the more 
recent opacity distribution functions of \cite{castellinum1}, but which 
could not remove this trend. A far more likely source 
for the trend is the limited accuracy of small 
log(gf)-values for weak lines offered in VALD. For example, 
we find that 14 Fe~{\sc i} and 7 Si~{\sc i} lines must 
be removed from the VALD line list because they are not 
observed in the Sun and Procyon. It points to problems with 
advanced calculations of semi-empriric (approximate) 
line oscillator strengths for complex model atoms 
of the iron-group elements. 
The accuracy of the predicted log(gf)-values for many lines 
of these complicated atoms is limited, and we find that 
they are systematically over-estimated for weak neutral lines.
An explanation for the over-estimated VALD log(gf)-values 
for our sample of Si~{\sc i} lines is less clear because 
many other Si~{\sc i} lines do correctly 
fit the solar spectrum using VALD data. 
We think however that it results from problems with 
a small number of multiplets from high energy levels 
($E_{\rm low}$$\simeq$5-6 eV) in 
the neutral Si atom, yielding oscillator strength 
values in VALD of low accuracy for both 
weak and medium-strong lines.

\section{Conclusions}

We measure the log(gf)-values of 1178 atomic absorption lines 
we identify in the Sun with advanced synthetic spectrum 
calculations. The new log(gf)-values of 660 lines are verified 
in Procyon. The measured log(gf)-values are available in the online 
SpectroWeb database and are corrections of the values provided 
in VALD. We find systematic over-estimations of 
the VALD values for many weak neutral lines of iron-group 
elements, and for a smaller sample of weak and medium-strong 
Si~{\sc i} lines. The log(gf) over-estimations are attributed 
to the limited accuracy of small log(gf)-values 
for weak lines presently offered in VALD. The oscillator strengths
of many weak and strong lines in both stars 
require further updates for reliable future line identifications.

\ack This work has been supported by the Belgian
Federal Science Policy - Terugkeermandaten.
Mr. P. Depoorter is gratefully acknowledged for 
assistance with the spectral line measurements.

\end{document}